\documentclass[a4paper,fleqn]{cas-dc}

\usepackage[sort, numbers]{natbib}

\usepackage{textgreek}
\usepackage{adjustbox}
\usepackage[shortcuts]{extdash}

\def\tsc#1{\csdef{#1}{\textsc{\lowercase{#1}}\xspace}}
\tsc{WGM}
\tsc{QE}
\tsc{EP}
\tsc{PMS}
\tsc{BEC}
\tsc{DE}

\begin{document}
\let\WriteBookmarks\relax
\def\floatpagepagefraction{1}
\def\textpagefraction{.001}
\shorttitle{Micropatterning photopolymerizable hydrogels for diffusion studies using pillar arrays or photomasks}
\shortauthors{Sevgi Onal et~al.}

\title [mode = title]{Micropatterning photopolymerizable hydrogels for diffusion studies using pillar arrays or photomasks}                      



\author[1]{Sevgi Onal}[orcid=0000-0002-9882-132X]
\cormark[1]
\ead{sevgional1@gmail.com}

\credit{Conceptualization, Methodology, Validation, Formal analysis, Investigation, Data curation, Visualization, Writing - Original Draft, Writing - Review \& Editing}

\address[1]{Istituto Italiano di Tecnologia, Naples, 80125, Italy}

\author[1,2]{Edmondo Battista} [orcid=0000-0003-0915-9176]

\credit{Methodology, Investigation, Writing - Review \& Editing}

\address[2]{Department of Engineering and Geology, University of G. d'Annunzio Chieti and Pescara, Pescara, 65127, Italy}

\author%
[1,3,4]
{Hilal Nasir} [orcid=0000-0003-4576-3872]
\credit{Methodology, Validation, Formal analysis, Investigation, Data curation, Visualization, Writing - Original Draft}

\address[3]{Interdisciplinary Research Center on Biomaterials (CRIB), University of Naples Federico II, Naples, 80125, Italy}

\address[4]{Clinical and Translational Oncology, Scuola Superiore Meridionale, University of Naples Federico II, Naples, 80138, Italy}

\author%
[1]
{Fabio Formiggini} [orcid=]
\credit{Methodology, Validation, Investigation, Visualization, Writing - Review \& Editing}

\author%
[1]
{Valentina Mollo} [orcid=]
\credit{Investigation, Writing - Review \& Editing}

\author[1]{Raffaele Vecchione}[orcid=]
\cormark[1]
\ead{raffaele.vecchione@iit.it}

\credit{Conceptualization, Methodology, Resources, Writing - Original Draft, Writing - Review \& Editing, Supervision, Project administration}

\author[1,3,5]
{Paolo Netti}[orcid=0000-0002-2435-7181]

\credit{Resources, Funding acquisition}

\address[5]{Chemical, Materials, and Industrial Production Engineering, University of Naples Federico II, Naples, 80125, Italy}


\cortext[cor1]{Corresponding authors}


\begin{abstract}
In situ polymerization and micropatterning of hydrogels on-chip opens the potential for many applications such as tracking and controlling the diffusion of molecules, stimulants, inhibitors, as well as nutrients and drugs, from their source to a target. To enable such applications, we developed hydrogel-on chip platforms for molecular diffusion studies by refining PEGDA-PEG hydrogel in terms of micropatterning and diffusion properties. In the first platform that we introduce here, the design has multiple adjacent microfluidic channels separated with pillar arrays shaping the flow of our custom-prepared photopolymerizable hydrogels and thus enabling the localization of photopolymerization. In the second platform, a photomask formation has been achieved by coupling the micro-milling of 250-\textmu m thickness of PMMA substrate with Platinum (Pt)-coating onto the PMMA mask. In this way the design was obtained to have opaque and transparent regions for light-based polymerization of the PEGDA-PEG hydrogels. The developed method of in situ polymerization of the hydrogel on-chip through photomask has enabled direct transfer of the design of interest to the pre-hydrogel in channel. Next, the developed platforms will be further used to test various compositions of photopolymerizable hydrogels and track and control the diffusion of molecules across hydrogel interfaces. The micropatterning methods and platforms developed here could be tailored to device design and development needs in various application fields from molecular transport to biosensing to electronic devices.


\end{abstract}





\maketitle

\section{Introduction}

With rapid progress in microfabrication techniques, micropatterning methods are being developed and used in a variety of applications, from materials engineering and on-chip technologies for healthcare to electronic devices \cite{ma2025photolithographic, zhu2024generation, mi13020164}. Microfabrication techniques such as photolithography, soft lithography, etching, rapid prototyping, micromachining, and thin film deposition are sharing common principles in various use cases, from microelectronics such as microbatteries and transistors within the semiconductor industry to microelectromechnical systems (MEMs) and microfluidics to microphysiological systems (MPS) and biosensors \cite{ma2025photolithographic, betancourt2006micro, kumar2024microfabrication}. The miniaturization achieved with these tools enables an abbreviation of long diffusion times, providing precise control over the materials, molecules, and processes to be investigated \cite{betancourt2006micro}.

Molecular transport tracking across the hydrogel and aqueous environments plays an important role in energy related biosensing and metabolic system. PEGDA hydrogels have attracted the attention of the scientific community to study and investigate molecular transport and movements because of their mesh size, hydrations, and functional group density, which influence how molecules such as metabolites diffuse and signal through soft materials. A recent study by Ding et al. (2025) investigated and described the role of hydrogel microstructures in regulating biomolecule transport, particularly focusing on energy-related analytes such as lactate, a key metabolic output \cite{ding2025comprehensive}. Herrmann et al. (2021) focus on how hydrogel porosity affects molecular flux and sensor responsiveness \cite{herrmann2021hydrogels}. According to Wu et al. (2024) \cite{wu2024interstitial} and Song et al. (2023) \cite{song2023flexible}, wearable and microfluidic hydrogel-based metabolic energy monitoring devices rely on this diffusion behavior to assess biomarkers of ATP production, exercise intensity, and oxidative stress. Furthermore, hydrogel membranes can modify transport rates to optimize signal output, as demonstrated by Nash et al. (2022) \cite{nash2022new}.

In situ polymerization and micropatterning of hydrogels on a chip enable investigation of various processes such as tracking and controlling the diffusion of molecules, stimulants, inhibitors, nutrients and drugs from their source to a target in various applications \cite{onal2021breast, son2017detecting, onal2024insitu, onal2025development}. To enable such applications, we developed hydrogel-on-chip platforms and light-based hydrogel polymerization methods that integrated the use of custom-prepared synthetic hydrogels \cite{onal2024insitu, onal2025development}. Naturally derived hydrogels such as Matrigel and collagen are widely used in assays, including on-chip platforms, but they possess a high batch-to-batch variation \cite{onal2021breast}. Alternatively, synthetic hydrogels offer more consistent properties, including tunable formulations, textures, and porosity. Light-based polymerized synthetic hydrogels provide precise control over these properties \cite{son2017detecting, onal2024insitu}. 

\begin{figure*}
\centering
  \includegraphics[width=1\textwidth]{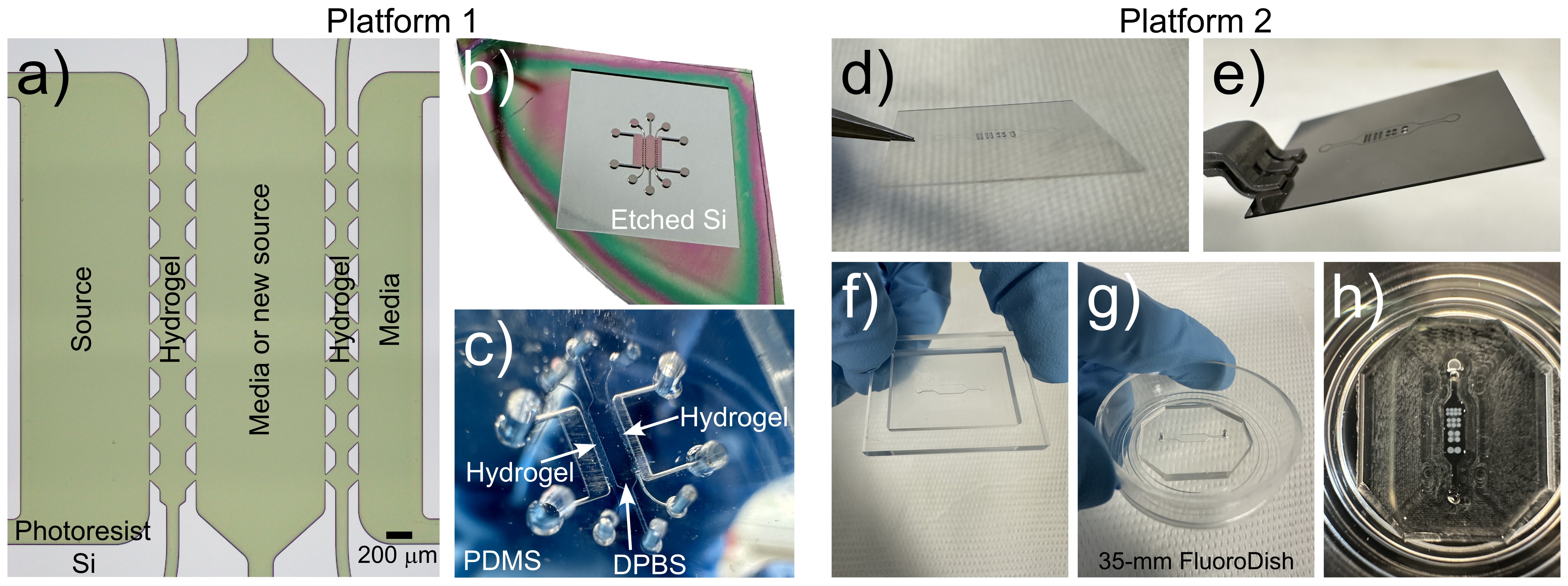}
  \caption{Micropatterning methods used to design and fabricate master and devices, represented in two platforms. (a-c) Hydrogel-on-chip device used for controlling diffusion of molecules to the adjacent channels through barriers of custom-prepared hydrogel. (a) Optical microscope stitched images of Si/photoresist master developed for the design written with laser lithography. (b) Etched Si master fabricated using ICP-RIE. (c) PDMS device with hydrogel and DPBS (Dulbecco's Phosphate-Buffered Saline) loaded into corresponding channels. (d) Microfabricated PMMA mask. (e) Pt-coated PMMA mask. (f) PMMA master of microchannel. (g) PDMS channel bonded to 35-mm Fluorodish glass. (h) PEGDA-PEG hydrogel cylinders in situ formed in PDMS channel through the Pt-coated PMMA mask.}
  \label{fig1}
\end{figure*}

The synthetic PEGDA-PEG hydrogels formulated in this work could be micropatterned using two platforms. The first platform incorporated adjacent microfluidic channels separated by pillar arrays having high-aspect-ratio ($\le5$) features, enabling the localized polymerization of custom photopolymerizable hydrogels (Figure~\ref{fig1}(a-c)). Fabrication involved silicon master production using laser lithography and inductively coupled plasma etching for deep silicon etching process. Polydimethylsiloxane chips were cast from these masters and used to characterize the diffusion of fluorescent molecules in different sizes, structures, and concentrations across channels. The second platform introduced a photomask-based method that includes micro-milling of 250-\textmu m-thick PMMA substrate and coating with platinum, which was aligned to a straight microchannel for the in situ formation of hydrogel cylinders on-chip (Figure~\ref{fig1}(d-h)). The photomask obtained with opaque and transparent regions was used to polymerize PEGDA-PEG hydrogels in microchannels using light exposure at wavelengths below 400-nm by a mercury lamp. This approach facilitated the rapid and customizable transfer of designs to pre-hydrogels, creating arrays of hydrogel cylinders within 200-\textmu m-high channels. In-situ formed hydrogel cylinders on-chip were further used to track the diffusion of individual molecules such as Rhodamine 110 and the chemotherapeutic drug Epirubicin hydrochloride, the diffusion and separation of mixed forms of Rhodamine 110 and Dextran-70-kDa, and the behavior of nanoparticles around the hydrogel cylinders. These examples demonstrated the ability of the platform to distinguish between small and large molecules. This platform has also been used to develop 3D antibody-based assays. Next, our platforms will be extended to study various photopolymerizable hydrogel compositions \cite{son2017detecting}, molecular diffusion and separation \cite{snyder2011experimental}, and cell-secreted molecular diffusion and capture for biosensing \cite{wang2025microfluidic, siavashy2024recent, goy2019microfluidics}. In order to distinguish between the terms used for our platforms, PDMS microstructures built in arrays separating the adjacent channels and mediating the hydrogel loading within the narrow channels are herein called pillars, while the hydrogel microstructures in situ formed inside a straight PDMS channel through a photomask are called cylinders. As presented, micropatterning in this work has been achieved not only by photolithography, which is a broad micropatterning technique, and etching techniques but also by micromilling and metal coatings. In this regard, the methods developed and used for representative applications here expand the capacities of micropatterning for device design and fabrication for a wide range of applications. 

\section{Materials and methods}
Here, we describe our microfabrication and surface activation methods with various coatings. The tools used to design and fabricate our chip platforms, the synthetic hydrogels formulated and micropatterned within these platforms, advanced imaging of molecular diffusion, gel structure, antibody immobilization in hydrogel cylinders, and image analysis are explained. We provide greater details for recreation of our platforms and reproducibility of our methods to be used in a wide range of applications in need of micropatterning. 

\subsection{Micropillar-based design and master microfabrication}
Fusion360 and AutoCAD (Autodesk) were used to create the designs. Silicon (Si) wafer was prebaked at 100\textsuperscript{o}C for 5 minutes. The AZ 4533 photoresist was spin-coated on a 2- or 4-inch Si wafer, employed 4-inch for fabrication of multiple designs and/or repeats of a design, at 4000 rpm for 1 minute with 500 rpm acceleration. The spin-coated photoresist/Si wafer was prebaked at 100\textsuperscript{o}C for 50 seconds. The conversion of the design file from .dxf to .job file was made using the so-called Conversion program. Then the design was written onto the AZ4533 photoresist (MicroChemicals GmbH) spin-coated wafer with direct write laser lithography (DWL 66 fs, Heidelberg Instruments). After laser writing, the master was post baked at 115\textsuperscript{o}C for 50 seconds. The photoresist master was developed with the AZ 351 B developer (MicroChemicals GmbH).

For deep Si etching, the Bosch process was applied using inductively coupled plasma reactive ion etching (ICP-RIE, Oxford Instruments). In the Bosch process, C\textsubscript{4}F\textsubscript{8} (for the deposition step) and SF\textsubscript{6} (for the etching step) gases were used for Si etching. After the Si etching recipe was complete, to finalize master fabrication, the etched Si master was treated for photoresist removal with AZ 100 Remover (MicroChemicals GmbH) applied for 10 min, washed with deionized water and dried with compressed air.

\subsection{PDMS device fabrication}
Silanization solution was prepared by mixing 94\% isopropanol, 4\% dH\textsubscript{2}O, 1\% acetic acid and 1\% Fluorolink to compose the working solution \cite{criscuolo2020double}. Before silanization, the etched Si master was treated with Oxygen plasma for 1 min at RF power of 50 W. The master was immersed in and coated with the silanization solution for 2 minutes. Then, the master was removed from the solution and placed at 90\textsuperscript{o}C for 1 hour. Alternatively, trichloro(1H,1H,2H,2H-perfluorooctyl)silane (Sigma-Aldrich) was used to coat the Si master surface by evaporation and deposition in a vacuum desiccator to easily remove the PDMS stamp after polymerization \cite{onal2021flexible}.
PDMS was prepared by mixing PDMS base and curing agent at a ratio of 10:1. After degassing in desiccator, PDMS was poured on the master and let to polymerize either for 2 days at room temperature or 1 day at room temperature and 2 hours at 80\textsuperscript{o}C. Then PDMS was cut and punched at inlets and outlets. For surface cleaning purposes, it was sonicated in H\textsubscript{2}O for 10 minutes and 70\% Ethanol for 5 minutes. After PDMS was dried with compressed air, it was also allowed to dry overnight at room temperature before device bonding. To complete the PDMS device fabrication, the PDMS stamp and glass slide were bonded immediately after surface treatment with Oxygen plasma for 30 seconds at RF power of 50~W. Finally, the devices were baked at 80\textsuperscript{o}C for at least 24 hours.

\subsection{Hydrogel preparation}
The synthetic hydrogel micropatterned within both platforms was formulated using PEGDA 700 Da 5\% and PEG 1000 Da 40\% (of the 80\% prepared stock), mixed with the final concentration of the photoinitiator Darocur 1\% and completed with deionized water.
Example for a total 2 ml hydrogel solution:
\begin{itemize}
\item	1 ml of 80\% PEG 1000 Da;
\item	100 \textmu l PEGDA 700 Da;
\item	20 \textmu l Darocur 1173 (2-Hydroxy-2-methylpropiophenone);
\item	880 \textmu l Deionized water.
\end{itemize}

In the modified hydrogel composition for fluorescent antibody conjugation experiments, acrylic acid was added 10\% of the PEGDA volume. 

Example for a total 2 ml hydrogel solution with acrylic acid:
\begin{itemize}
\item	1 ml of 80\% PEG 1000 Da;
\item	100 \textmu l PEGDA 700 Da;
\item	20 \textmu l Darocur 1173 (2-Hydroxy-2-methylpropiophenone);
\item	10 \textmu l Acrylic acid;
\item	870 \textmu l Deionized water.
\end{itemize}

\subsection{PDMS pillar array-mediated chip loading with hydrogel and light-based polymerization}
The chips were loaded with hydrogels in the narrow channels confined with pillar arrays with a manual hand pipette and photo-polymerized for 30 seconds. The custom-modified microscope setup using light exposure at wavelengths below 400 nm by high intensity mercury lamp (Lumen Dynamics X-Cite 120PC Q) at 100\% power (160 mW) with a 2X objective was used to polymerize PEGDA-PEG hydrogels.

\subsection{Microfluidic device fabrication for in situ hydrogel cylinder formation}
The PMMA master and photomask were fabricated using micromilling through 3-mm and 250-\textmu m PMMA sheets, respectively. PDMS prepared in a ratio of 10:1 PDMS base and curing agent was poured onto the 3-mm-thick PMMA master and allowed to polymerize for 2 hours at 80\textsuperscript{o}C after degassing in the desiccator.

PDMS molds were cut to a size that can fit within a 35-mm FluoroDish. Inlets/outlets were punched with a 1.2 mm puncher. PDMS and FluoroDish glass were bonded using a plasma treatment for 30 seconds at RF power of 50 W. The devices were heated at 80\textsuperscript{o}C for 1 hour after bonding. 

\subsection{Surface modification of the microchannels}
To make the surfaces appropriate for hydrogel patterning in microchannels, we treated the surfaces for 30 seconds with a solution of 2\% 3-(Trimethoxysily)propyl methacrylate (Sigma-Aldrich) in acetone. The microchannels were washed with dH\textsubscript{2}O 3 times for 3 min/wash and then heated at 80\textsuperscript{o}C for 4 hours for drying.

\subsection{Chip loading with hydrogels}
Straight microchannels were loaded with the hydrogels that were prepared with or without acrylic acid, with a manual hand pipette. The channel was then aligned with our custom-made photomask to allow in situ patterning of the photopolymerizable hydrogel inside. The same custom-modified microscope used with the PDMS pillar array-mediated chip was utilized here at 12\% power (15 mW) with a 2X objective to polymerize PEGDA-PEG hydrogels for 30 seconds. After light exposure, the unpolymerized gel was washed with DPBS 3 times, revealing the hydrogel cylinders formed inside microchannels.
The chips were then stored in the refrigerator for at least 2 days, allowing the hydrogel structures to be statically washed by changing DPBS every day.

\subsection{Chip loading with molecules, LSCM imaging and analysis}
After 2 days of static washing of the hydrogels within the chips stored at 4\textsuperscript{o}C, microchannels were washed with DPBS three times. Meanwhile, microscope incubation conditions were set at 5\% CO\textsubscript{2} and 37\textsuperscript{o}C. After the chip was loaded with a certain concentration of each type of molecule, imaging was immediately started using a Laser Scanning Confocal Microscope (LSCM; Leica Microsystems, Germany) to capture molecular diffusion from the zero time of the loading.
The 10X objective (HCX APO L U-V-I 10.0x0.30 WATER or HCX FL PLAN 10.0x0.25 DRY) was used to capture diffusion across microchannels and hydrogel lines in the pillar-based chip and hydrogel cylinders in the straight microchannel. For higher mignification imaging of the gel structure on-chip, the 25x objective (HCX IRAPO L 25.0x0.95 WATER) was used at reflection mode.

Image analysis was performed to quantify the diffusion of molecules from the source channel (where they were introduced) to the hydrogel and DPBS channels on 12-bit images using ImageJ 1.54p (Fiji) \cite{schindelin2012fiji}. Regions of interest (ROIs) were drawn in respective regions following the design in brightfield channels and correlated to the same regions in fluorescent channels as shown in Supplementary Figure~\ref{SuppFig1}(a). The intensity of the fluorescence signal was then measured and plotted as a normalized mean gray value along these ROIs in different chips for different molecules, such as BSA and dextran (Figure~\ref{fig2} and Supplementary Figure~\ref{SuppFig2}). For different concentrations of the same molecule type, the signal was measured directly as the gray value along a plot profile (Figure~\ref{fig3}). During these image acquisition and analysis, the focal plane in which the images were taken for the quantification of the fluorescence intensity was the central plane of the z-volume taken from the bottom (glass surface) to the top (PDMS surface) of the channel, as represented in Supplementary Figure~\ref{SuppFig1}(b).

\subsection{SEM imaging}
Scanning electron microscopy (SEM) imaging was performed on a polymerized gel within the microchannel confined with pillar arrays, as well as on the PDMS pillars themselves, and on hydrogel cylinders in situ formed inside a microchannel. The bonded PMDS mold was cut at the edges of microchannels to reach the inside patterns under cryogenic conditions using liquid nitrogen. The patterns were mounted on aluminum pin stubs using conductive carbon tape and sputter-coated with a 15 nm gold layer (Sputter coater HR208, Cressington) prior to SEM acquisition (Ultraplus, Zeiss). The imaging was performed at an accelerating voltage of 5 kV, with magnifications ranging from 60x to 70kx using a secondary electron detector.

\subsection{Antibody immobilization in hydrogel cylinders on-chip}

\subsubsection{Surface activation}
A freshly made solution containing 0.20 M EDC (C\textsubscript{4}H\textsubscript{1}N\textsubscript{3}HCl) and 0.40 M NHS (C\textsubscript{4}H\textsubscript{5}NO\textsubscript{3}S) in 0.1 M MES buffer (pH 5.5) was added to the microfluidic channel and allowed to sit at room temperature for 30 minutes. Surface carboxyl groups are transformed into reactive NHS-esters by this carbodiimide-mediated process, which can then pair with the main amine groups of biomolecules. Conjugation efficiency is increased by maintaining an acidic, amine-free buffer such as MES, which reduces side reactions and increases NHS-ester stability. The fundamental mechanism of carbodiimide chemistry is presented by Hermanson (2013), who explains how EDC activates carboxyl groups to make O-acylisourea intermediates, which are then stabilized by NHS to produce amine-reactive esters \cite{hermanson2013bioconjugate}. All reagents, buffers, and antibodies were obtained from Sigma-Aldrich (USA).

\subsubsection{Removal of activation reagents}
After activation, the channel was rinsed three times with MES buffer for five minutes at intervals of five minutes to remove any remaining coupling chemicals and urea byproducts.

\subsubsection{Primary antibody immobilization}
Primary antibody Rabbit IgG diluted in PBS (pH 7.4) was applied to two experimental chips to final concentrations of 10 \textmu g ml\textsuperscript{-1} and 1 \textmu g ml\textsuperscript{-1}, respectively. As a negative control, a third chip was just given PBS. To promote the formation of amide bonds between NHS-esters and antibody amine groups, solutions were statically incubated for 2.5 hours at room temperature. The 2.5-hour incubation period was chosen to minimize antibody denaturation and guarantee effective coupling.

\subsubsection{Quenching of unreacted NHS-esters}
The channels were rinsed three times with 0.1M Tris buffer (pH 7.4) to hydrolyze any remaining active esters and prevent nonspecific binding.

\subsubsection{Equilibration}
Subsequent washing with DPBS (three times, pH 7.4) removed unbound antibodies and equilibrated the microenvironment for labeling.

\subsubsection{Secondary antibody labelling}
To guarantee consistent diffusion and binding within the hydrogel matrix, secondary fluorescent Cy5 anti Rabbit IgG (1.2 \textmu g ml\textsuperscript{-1} in PBS) was added and incubated for approximately 16 hours at 4\textsuperscript{o}C in the dark.

\subsubsection{Final washing and imaging}
After overnight conjugation, chips were rinsed three times with DPBS (5 min each) and immediately imaged using LSCM. Fluorescence imaging of antibody-functionalized hydrogel microstructures was performed using a Leica SP5 II Laser Scanning Confocal Microscope (Leica Microsystems, Germany). Samples were imaged using an HCX APO L U-V-I 10×/0.30 water-immersion objective with scan field 1.55 × 1.55 mm, frame size 512 × 512 pixels and in z-stack mode a total of 51 z-sections were collected with a step size of 3.21 \textmu m and a pinhole diameter set to the optimized default value of 1 A.U. (Airy Unit), corresponding to voxel dimensions of  3.03 \textmu m × 3.03 \textmu m × 3.21 \textmu m. Two channels were simultaneously acquired by PMT (PhotoMultiplier Tubes) detectors: Cy5 far-red fluorescence and brightfield. Cy5 excitation was provided by a He-Ne 633 nm laser, and emission detection was set in the 612-723 nm range. The scan speed was set to 400 Hz, without averaging (to ensure faster acquisition). Images were acquired at 12-bit resolution. All parameters were kept constant across the samples to ensure quantitative comparability.

\subsubsection{Image analysis}
All imaging data were analyzed using the latest version of ImageJ 1.54p (Fiji) \cite{schindelin2012fiji}. Each microfluidic chip contained 20 hydrogel cylinders of varying diameters. For quantitative analysis, cylinders of the same diameter were selected, and a region of interest (ROI) was drawn for each. Fluorescence intensity was measured within each ROI and the mean gray value was calculated for all 20 cylinders to determine the average signal intensity corresponding to fluorescent labeling.

\section{Results and discussion}

We developed hydrogel-on-chip platforms for molecular diffusion studies by refining PEGDA-PEG hydrogel in terms of micropatterning and diffusion properties. We used two different methods to create masters and microdevices for various applications where micropatterning is paramount. The first method is by coupling direct write laser lithography with inductively coupled plasma reactive ion etching (ICP-RIE) to transfer the intended design to a Si master and shape the loaded hydrogels through design of adjacent channels separated by pillar arrays. This approach was selected to create narrow channels with pillar arrays having high-aspect-ratio ($\le5$) features, ensuring durability and high-quality surface finishes on the Si masters.

The second method relies on aligning a photomask to a microchannel to shape the photopolymerizable hydrogel into cylinders by light passing through the holes in the photomask that is microfabricated using micromilling and metal coating. Thus, in the platform introduced here, a photomask formation has also been achieved by coupling the micro-milling of 250-\textmu m thickness of PMMA substrate with Platinum (Pt)-coating onto the PMMA mask. In this way, the design was obtained to have opaque and transparent regions for light-based polymerization of the PEGDA-PEG hydrogels. 

To characterize the thickness of the photoresist and then the height of etched Silicon (Si) master for casting PDMS pillars and channels, a Dektak stylus profilometer was used. The photoresist height processed by direct write laser lithography on the Si wafer (Figure~\ref{fig1}(a)) was measured to be 3.40~\textmu m. The feature heights of the etched Si master (Figure~\ref{fig1}(b)) were measured to be 198.5 \textmu m and 197.8 \textmu m around the channels after etching and after removal of photoresist, respectively.

\begin{figure*}
 \centering
 \includegraphics[width=\textwidth]{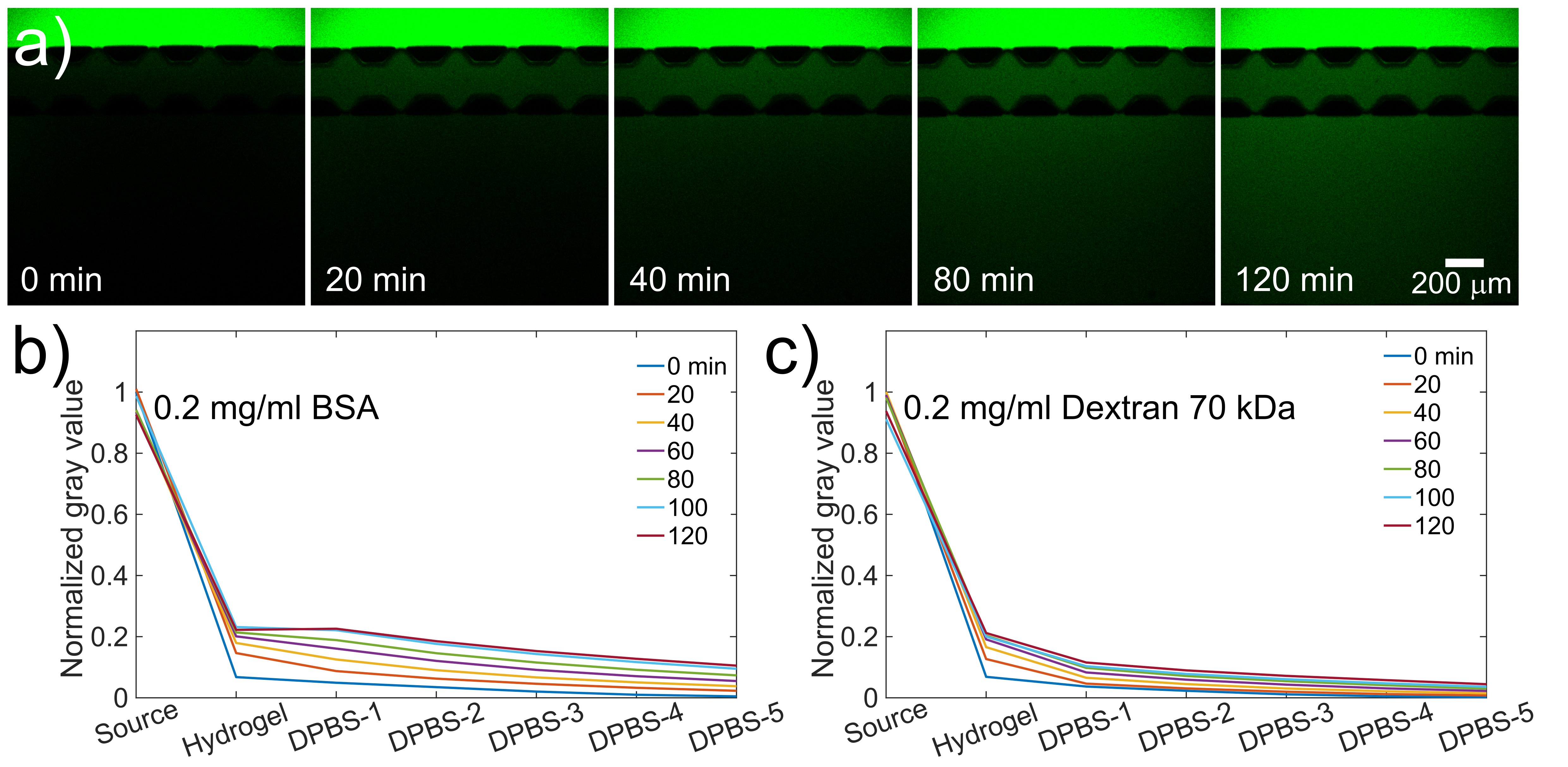}
 \caption{LSCM images showing diffusion of fluorescent molecules and quantification. (a) Time lapse images showing diffusion of Bovine Serum Albumine (BSA) fluorescein across channels from source channel (top) to hydrogel in narrow channel between pillar arrays and to DPBS channel (bottom) over 2 hours, shown here as representative data. (b-c) Quantification of molecular diffusion from source channel to hydrogel and to DPBS channel for BSA (b) versus Dextran 70 kDa (c). Channel regions and ROIs used in quantification are shown in Supplementary Figure~\ref{SuppFig1}(a).}
 \label{fig2}
\end{figure*}

\begin{figure*}
 \centering
 \includegraphics[width=1\textwidth]{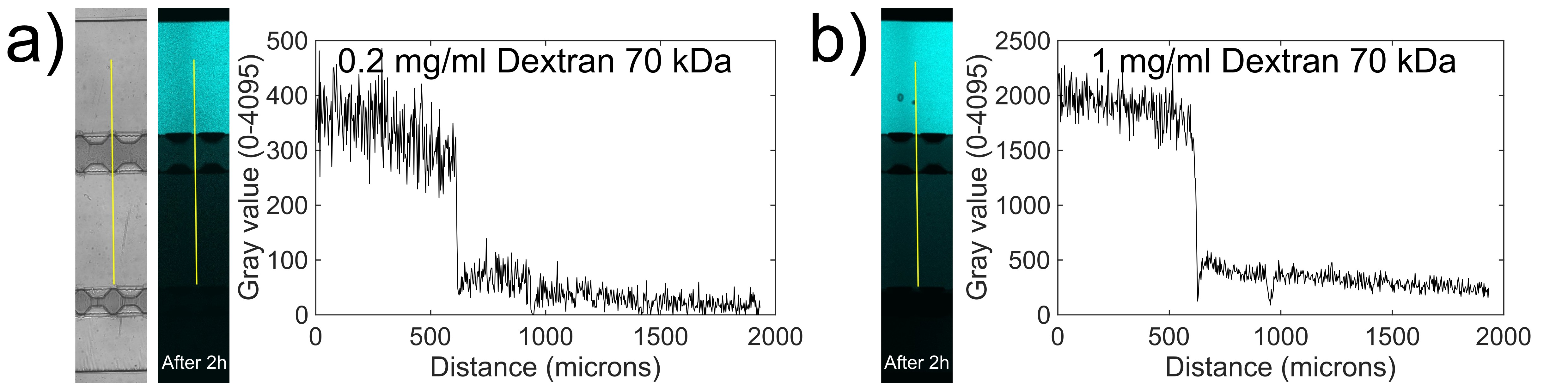}
 \caption{LSCM imaging and quantification of molecular diffusion from source channel to hydrogel and to DPBS channel by tile scanning and comparison of Dextran 70 kDa concentrations of (a) 0.2 mg/ml and (b) 1 mg/ml in DPBS for diffusion length.}
\label{fig3}
\end{figure*}

We synthesized and loaded PEGDA based hydrogels with PEG as porogen into the chips (Figure~\ref{fig1}(c) and Supplementary Figure~\ref{SuppFig1}). Gel polymerization was performed under a microscope using light exposure at wavelengths below 400~nm by high intensity mercury lamp. Fluorescent molecules, including BSA conjugates and dextrans of varying molecular weights, were loaded into the corresponding channels (Figure~\ref{fig2}(a) and Supplementary Figure~\ref{SuppFig2}(a)). We conducted 3D scans (z-stacks) using confocal microscopy (Supplementary Figure~\ref{SuppFig1}(b)) and quantified molecular diffusion through time-lapse (Figure~\ref{fig2}) and tile scan (Figure~\ref{fig3}) imaging. Our observations included differences in molecular diffusion of BSA and Dextran 70 kDa (Figure~\ref{fig2}(b-c)), likely due to the molecular structures of BSA \cite{singh2020aqueous} versus dextran \cite{diaz2021dextran} for being globular versus branched, respectively, and the resulting porosity of the PEG porogen (1000 Da) within the PEGDA hydrogel (700 Da). We also observed variations in diffusion rates among dextrans of different molecular weights (Supplementary Figure~\ref{SuppFig2}(b)). Furthermore, we investigated the effect of the initial concentration of the molecules, e.g., 0.2~mg/ml versus 1 mg/ml of Dextran 70 kDa, on the length of their diffusion across the channels, as illustrated in Figure~\ref{fig3}, as well as between 0.1 mg/ml and 1 mg/ml of Dextran 40 kDa or Dextran 10 kDa (data not shown).

\begin{figure*}
 \centering
 \includegraphics[width=0.9\textwidth]{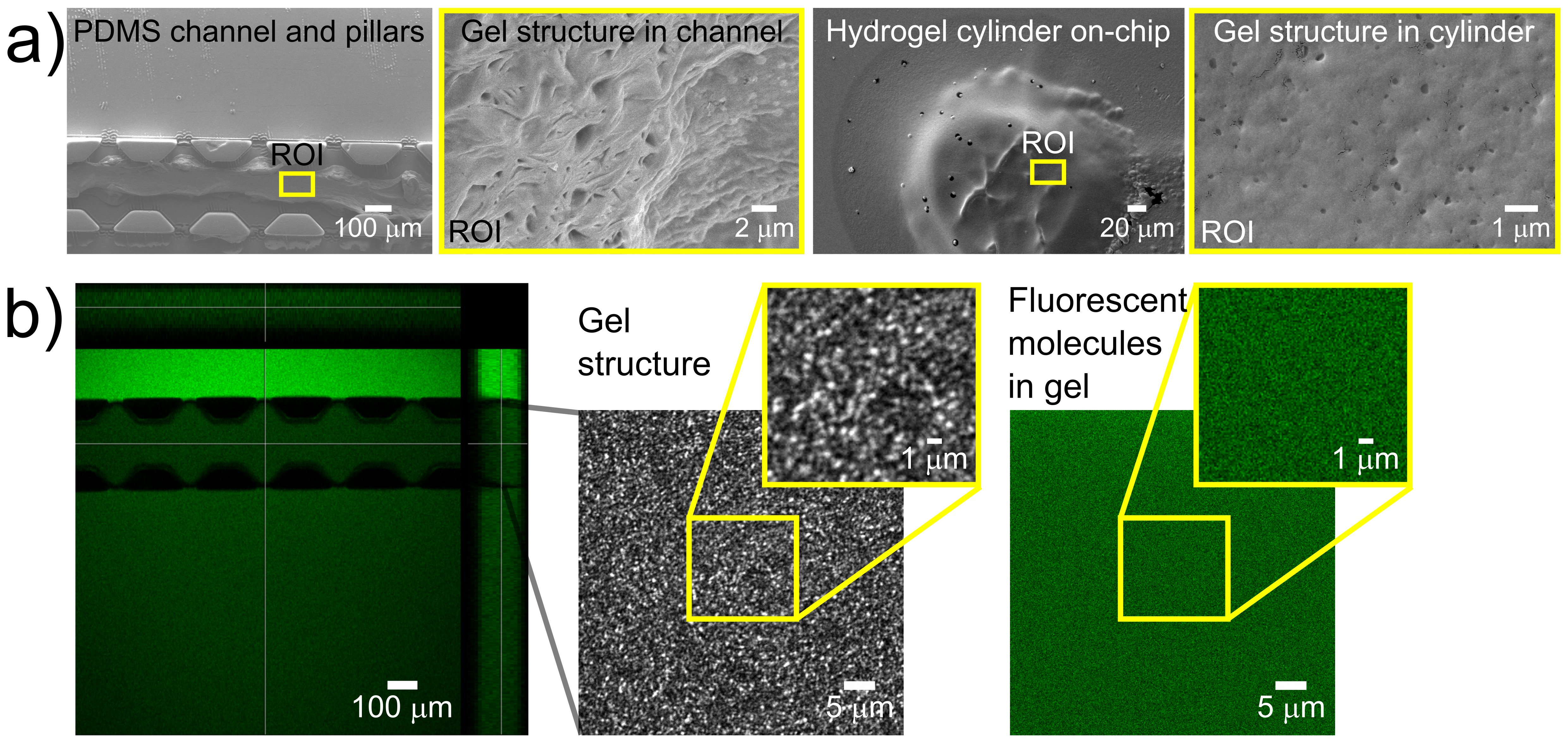}
 \caption{Imaging of hydrogel-on-chip structures. (a) SEM images of PEGDA-PEG hydrogel in the chip with PDMS pillars and in the chip with hydrogel cylinders, formed by polymerizing at 100\% power (160 mW) and 12\% power (15 mW), respectively. (b) Representative LSCM image (z-stack) of a chip with orthogonal projections. LSCM images of the hydrogel structure by reflection and the molecules inside gel by fluorescence mode.} 
  \label{fig4}
\end{figure*}

The gel structure was investigated using SEM and laser-scanning confocal microscopy. PDMS pillar arrays and the hydrogel in between, as well as the structure within the hydrogel cylinders in situ formed in a straight PDMS channel (Figure~\ref{fig4}(a)). Furthermore, the reflection mode of laser-scanning confocal microscopy shows the structure of the hydrogel within the pillar arrays, while the fluorescence mode shows the fluorescent molecules distributed in this hydrogel structure (Figure~\ref{fig4}(b)). 

The method developed for in situ polymerization of the hydrogel on-chip through a photomask has enabled direct transfer of the design of interest to the pre-hydrogel in a 200-\textmu m-height channel (Figure~\ref{fig1}(f-h) and Supplementary Figure~\ref{SuppFig3}(a)), also comparable to the polymerization in a bulk sample created as a hydrogel strip with a pipette (Supplementary Figure~\ref{SuppFig3}(b)). Even when a better formation of the hyrogel cylinders was achieved in a bulk sample, they collapsed during washing steps with DPBS removing the unpolymerized gel, as illustrated in Supplementary Figure~\ref{SuppFig3}(c), due to the absence of a surrounding microchannel to confine them. Thus, the method is relatively faster to change the design of interest shaped into a final array of hydrogel cylinders (Figure~\ref{fig1}(h) and Supplementary Figure~\ref{SuppFig4}). In these applications, the mercury lamp was set at 12\% power instead of 100\% to minimize light scattering. 

\begin{figure*}
 \centering
 \includegraphics[width=1\textwidth]{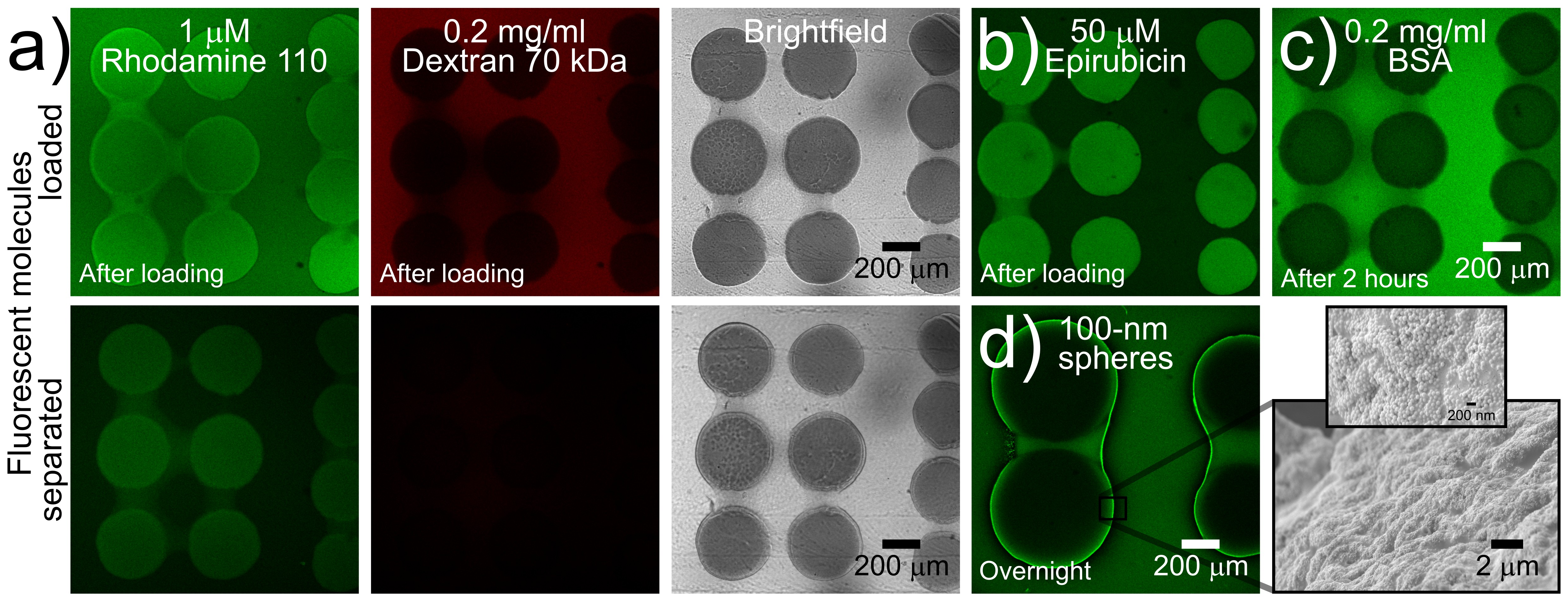}
 \caption{Diffusion of various molecules in hydrogel cylinders on-chip. (a) Diffusion and separation of a mixed form of Rhodamine 110 and Dextran 70 kDa. (b-c) Diffusion of Epirubicin hydrochloride and BSA into the gel cylinders. (d) The behavior of 100-nm spheres around the gel, by LSCM and SEM.} 
  \label{fig5}
\end{figure*}

The in situ formed hydrogel cylinders on-chip were further employed to track the diffusion of individual small molecules such as Rhodamine 110 (Supplementary Figure~\ref{SuppFig4}(a-b)), diffusion and separation of mixed forms of Rhodamine 110 and Dextran-70-kDa (Figure~\ref{fig5}(a)) as different size molecules, diffusion of chemotherapeutic drug Epirubicin hydrochloride (Figure~\ref{fig5}(b) and Supplementary Figure~\ref{SuppFig4}(c)), fluorescein conjugated BSA (Figure~\ref{fig5}(c)), and the behavior of nanoparticles around hydrogel cylinders (Figure~\ref{fig5}(d)). The 3D formation of the hydrogel cylinders within the microchannel was represented by a 3D scan (z-stack) of the hydrogel cylinders with the diffusion of Epirubicin hydrochloride, shown in the 3D view in Supplementary Figure~\ref{SuppFig4}(c).

\begin{figure*}
 \centering
 \includegraphics[width=1\textwidth]{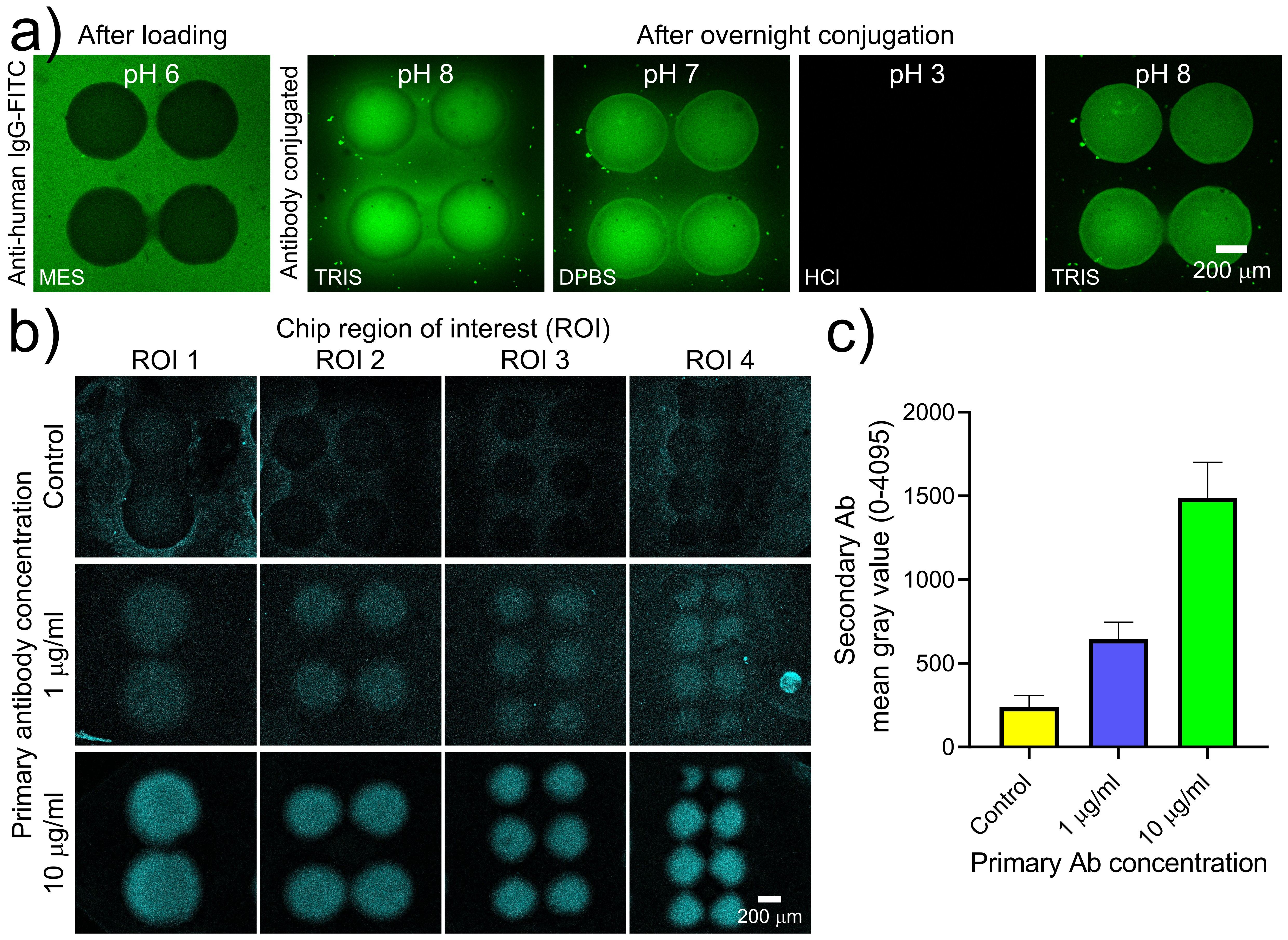}
 \caption{Antibody conjugation and capture in 3D using hydrogel cylinders on-chip. (a) Fluorescent antibody conjugation to the gel. (b) Fluorescent secondary antibody captured by primary antibody conjugated to the hydrogel in different concentrations including 1 \textmu g/ml and 10 \textmu g/ml. In control group, hydrogel cylinders were not conjugated with primary antibody. (c) Quantification of the secondary antibody fluorescent signal readout (mean gray value) as a function of the primary antibody concentration, after overnight interaction of primary and secondary antibodies in hydrogel cylinders.} 
  \label{fig6}
\end{figure*}

The diffusion and immobilization of the primary antibody within PEGDA-PED-acrylic acid hydrogel cylinders on-chip from after loading to after overnight conjugation were illustrated by first washing with Tris buffer (pH 8) and DPBS (pH 7) and quenching FITC (fluorescein isothiocyanate) of the fluorescent primary antibody (anti-human IgG-FITC) with HCl solution (pH 3) and then reactivating the signal with Tris buffer as the basic solution (pH 8) (Figure~\ref{fig6}(a)). These data have provided the basis for the next experiments with conjugation of non-fluorescent primary antibody into the gel structure and its interaction with fluorescent secondary antibody as detailed in Figure~\ref{fig6}(b-c).

Thanks to EDC/NHS chemistry \cite{hermanson2013bioconjugate}, primary and secondary antibodies were covalently attached to the surface of the 3D microenvironment of our PEGDA 700 Da hydrogel. The intensity of the signal increased proportionally to the concentrations of the primary immobilized antibody, as shown in Figure~\ref{fig6}(b). The primary antibody concentration was tested from lower to higher, as 1 \textmu g/ml was lower and 10 \textmu g/ml was the highest concentration of primary antibody immobilized through the gel surface, providing the quantification as shown in Figure~\ref{fig6}(c). The intensity of the fluorescence signal measured in mean gray value increased as the concentrations of the primary antibody increased.

Next, our custom-developed photomask-based method will be used to test various compositions of light responsive hydrogels \cite{son2017detecting}. Such new hydrogels will be further employed to track the diffusion of molecules across hydrogel interfaces, as presented here for the pillar array-mediated platform and in situ formed hydrogel cylinders on-chip.

The development and use of the pillar array-mediated chip platform for micropatterning hydrogels included several key design and fabrication optimizations. The use of trichloro(1H,1H,2H,2H-perfluorooctyl)silane (Sigma-Aldrich) by evaporation and deposition to coat the surface of the Si master with high-aspect-ratio features has better enabled easy and complete peeling of the PDMS stamp with intact pillar arrays after polymerization, compared to Fluorolink solution which is an aqueous-based coating and may not penetrate all the time into small cavities within the design of Si master. Thus, the silane evaporation and deposition-based coating increased the lifetime of the Si master by leaving the cavities PDMS remnant free to be repeatedly used for new PDMS casting. As shown in Figure~\ref{fig1}(c) and Supplementary Figure~\ref{SuppFig1}, in PDMS devices bonded to glass, the hydrogel flow ran properly in the respective channel in the presence of the intact pillar array adjacent to the side channel. Those were empty channels. Thus, the shape, angle, and spacing of the pillars, as well as surface hydrophobicity, helped direct and keep the flow running properly. 

In the photomask-based chip platform for in situ formation of hydrogel cylinders, micropatterning of the hydrogel in microfluidic settings outperformed the bulk hydrogel strip method, despite the fact that our custom-made photomask could be utilized in both conditions (Supplementary Figure~\ref{SuppFig3}). Furthermore, by coating the microchannel surface with 3-(Trimethoxysily)propyl methacrylate, the PEGDA-PEG hydrogel cylinders formed in situ adhered to the bottom and top sides of the microchannel were stable and durable during multiple solution loading and washing steps (Supplementary Figure~\ref{SuppFig4}(c)). The confinement of polymerized cylinders within a microchannel was developed as a contrast approach to stop-flow lithography \cite{dendukuri2007stop}.

Although SEM and LSCM images gave a representation of the gel structures in both platforms, a more detailed investigation of the pore size will give further insight into the structural formation of these gels. For the effect of molecular structure size controlling the passage of larger particles (e.g. 100-nm) into hydrogel cylinders, the results obtained with both imaging techniques are in agreement with each other, as shown in Figure~\ref{fig5}(d).

The diffusion examples shown in this work demonstrated the ability of our hydrogel-on-chip platforms to distinguish between molecules of similar size but with different structures, between different concentrations of the same molecules, and among molecules of different sizes. 

Compared with the hydrogel formulation used for the diffusion experiments of various molecules, the addition of acrylic acid in the relevant formulation created carboxyl sites for the primary antibody to attach. In this way, the primary antibody gets immobilized and then binds to the fluorescent secondary antibody that shows the signal readout in hydrogel cylinders. Di Natale \textit{et al.} (2021) confirmed that mildly acidic conditions improve coupling yield and reproducibility in on-chip systems by demonstrating a comparable EDC/NHS activation method on PDMS–PAA microfluidic interfaces for peptide immobilization \cite{di2021easy}. 

In agreement with the results in Figure~\ref{fig6}, a comparable concentration-dependent fluorescence enhancement has been documented in PEG-based hydrogel biosensing studies; a higher probe density increases the number of accessible binding sites, which enhances capture efficiency. Higher functional group density in hydrogels directly strengthens optical biosignals, as explained by Herrmann \textit{et al.} (2021) \cite{herrmann2021hydrogels}. Fluorescence signals in PEG/PEGDA matrices increase with probe loading, as shown by Mazzarotta \textit{et al.} (2021) \cite{mazzarotta2021hydrogel} and later verified by Sun \textit{et al.} (2024) \cite{sun2024recent} as well. Shikha \textit{et al.} (2018) show the same trend in hydrogel microbeads \cite{shikha2018upconversion}, and Nash \textit{et al.} (2022) highlight that hydrogel membrane structure modulates transport and signal strength \cite{nash2022new}. These findings collectively demonstrate that our PEGDA hydrogel structure facilitates the efficient diffusion of antibodies, or protein molecules of large and small size, and results in a predictable increase in fluorescence with higher concentrations of primary antibodies. 
The fluorescence response, which varies with concentration in our system, indicates that transport is effective at all tested levels of antibody loading, as demonstrated in Figure~\ref{fig6}(b-c). Such a varying concentration dependent diffusion and immobilization, and hence fluorescent signal readout can be obtained by multiple ROIs of hydrogel simultaneously tested on a chip thanks to our developed micropatterning methods. Thus, we showed the suitability of our methods for biosensing applications, while these methods can be used for designs for other sensing applications. Consequently, PEGDA hydrogels are well-suited for bio-energy applications that require controlled molecular transport, such as real-time metabolite sensing or hydrogel-integrated microfluidic energy monitoring platforms.

The molecular structure, size, and concentration-dependent signal readouts demonstrated with our platforms could be further implemented for energy applications tracking the transport of molecules across aqueous and gel environments. Our methods have been described in detail for their reproducibility across various fields. The different applications in this work have further shown the reproducibility of our micropatterning methods developed using different microfabrication technologies. 

On the other hand, to compare the two developed micropatterning methods for sustainability, the PMMA photomask-based master fabrication method using micromilling can be an eco-friendly alternative to the Si master fabrication method which is more energy-consuming to use direct write laser lithography and ICP-RIE that also uses green house gasses such as C\textsubscript{4}F\textsubscript{8} and SF\textsubscript{6} during the deep Si etching process \cite{nagano2007environment, wang2025green}. Thus, the scalability of our methods and platforms requires consideration of energy consumption and green house effect by the latter advanced microfabrication techniques, with a compromise on micropatterning resolution when the micromilling-based method is used.

Although they were fabricated using different technologies, the two chip platforms have a common underlying principle in micropatterning of photopolymerizable hydrogels using our microscope setup reconfigured for controlled photopolymerization through microdevices. Thus, the two platforms could enable the same applications in which the synthetic hydrogels could be micropatterned and utilized for molecular level studies. While the first platform with adjacent microchannels does not rely on a photomask and instead uses the PDMS pillar arrays to confine and shape the hydrogel, the second platform requires alignment of a microchannel to a photomask to transfer the design of interest to the hdyrogel and in situ form cylinders inside a microchannel. Having the ability to use the photomask repeatedly with many straight microchannels, the photomask-based platform could be perceived to be faster in transferring the design of interest and micropattern various compositions of hydrogels, thus helping the hydrogel optimizations for the PDMS pillar arrays-mediated platform as well.

\section{Conclusions}

The versatility of micropatterning photopolymerizable hydrogels in our developed chip platforms enabled various applications. The two platforms, despite being created using different microfabrication technologies, complement each other for molecular diffusion characterizations and applications. The selection of the platform and subsequent modifications in the hydrogel structure were based on the custom requirements of molecular diffusion, immobilization, and capture applications. The microdevice design and utilization approaches, micropatterning methods, and platforms developed in this work can be implemented in various applications, from molecular transport tracking to biosensing and on-chip technologies for healthcare and energy applications. 

\printcredits

\section*{Declaration of Competing Interest}
There are no conflicts of interest to declare.

\section*{Acknowledgements}
The authors thank Denise Pagliara, Carlo Natale, Pasquale Caruso and Raffaella Della Moglie for technical help, and Claudia Mazio for providing Epirubicin hydrochloride. This research was funded by Istituto Italiano di Tecnologia (Italy).

\section*{Data availability statement}
Data will be made available on request.

\bibliographystyle{cas-model2-names}

\bibliography{cas-refs}

\section*{Appendix A. Supplementary data}
Supplementary material related to this article can be found in the following.

\setcounter{figure}{0}
\renewcommand{\thefigure}{S\arabic{figure}}

\begin{figure*}
 \centering
 \includegraphics[width=1\textwidth]{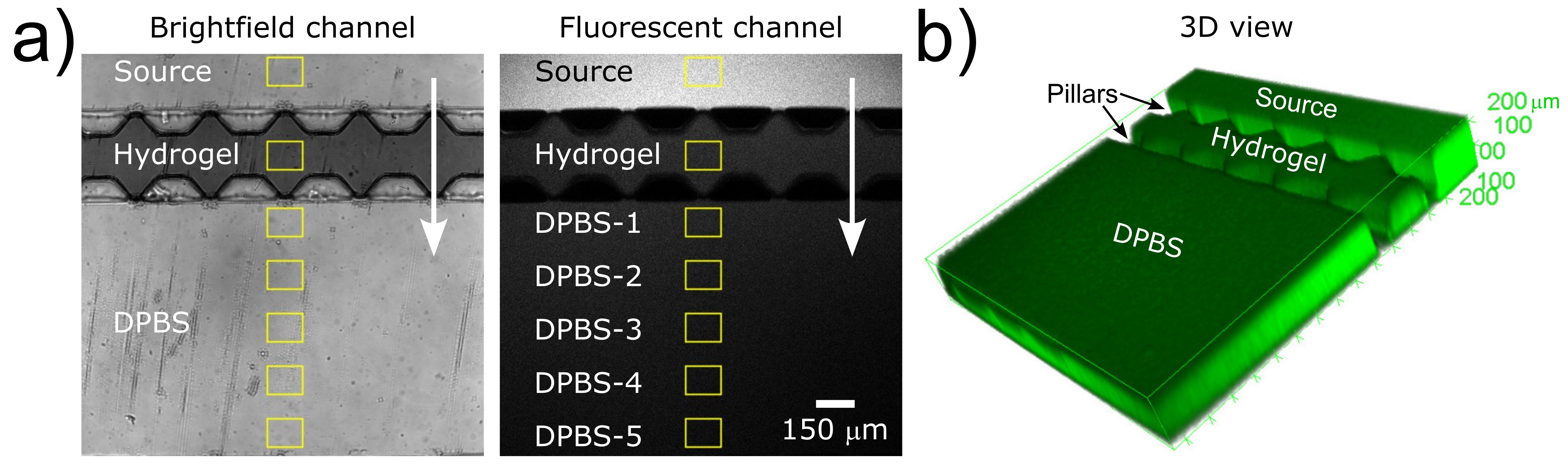}
 \caption{Image analysis and 3D imaging of the diffusion profile across microchannels. (a) ROI selection on brightfield and fluorescent images for image signal analysis. Arrows show the diffusion direction from source to hydrogel to DPBS channel. (b) LSCM image with 3D view.} 
  \label{SuppFig1}
\end{figure*}

\begin{figure*}
 \centering
 \includegraphics[width=1\textwidth]{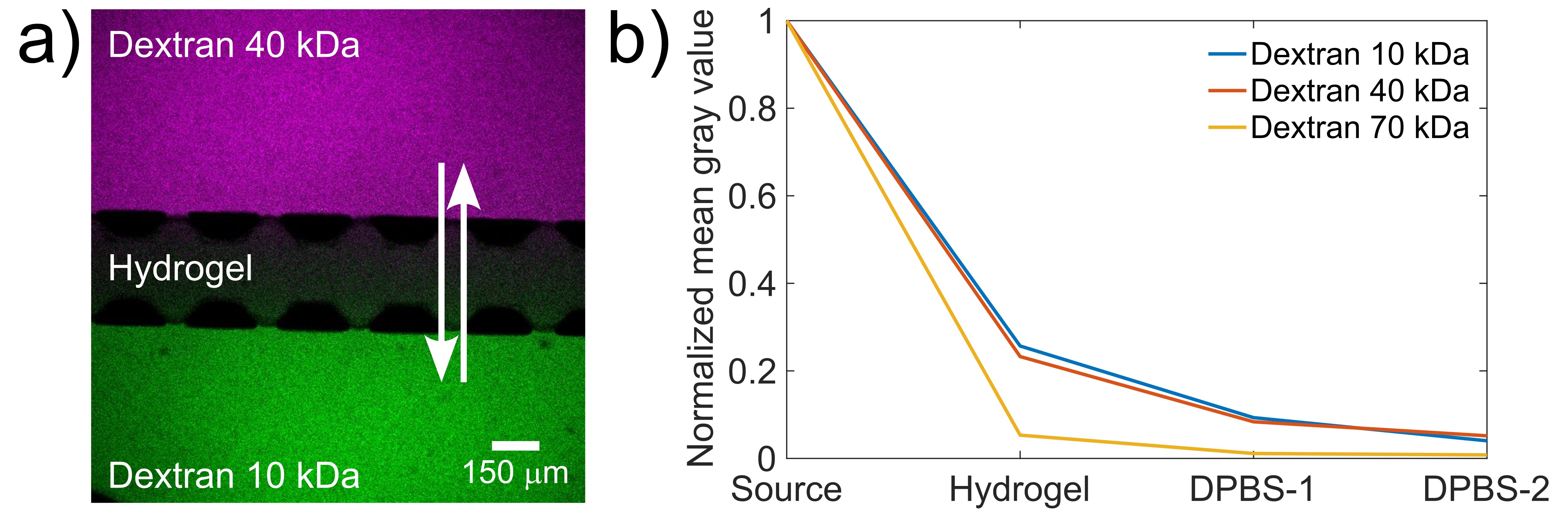}
 \caption{Diffusion comparison of different size molecules on-chip. (a) LSCM images and (b) quantification of diffusion of Dextran types with different molecular weights.} 
  \label{SuppFig2}
\end{figure*}

\begin{figure*}
 \centering
 \includegraphics[width=1\textwidth]{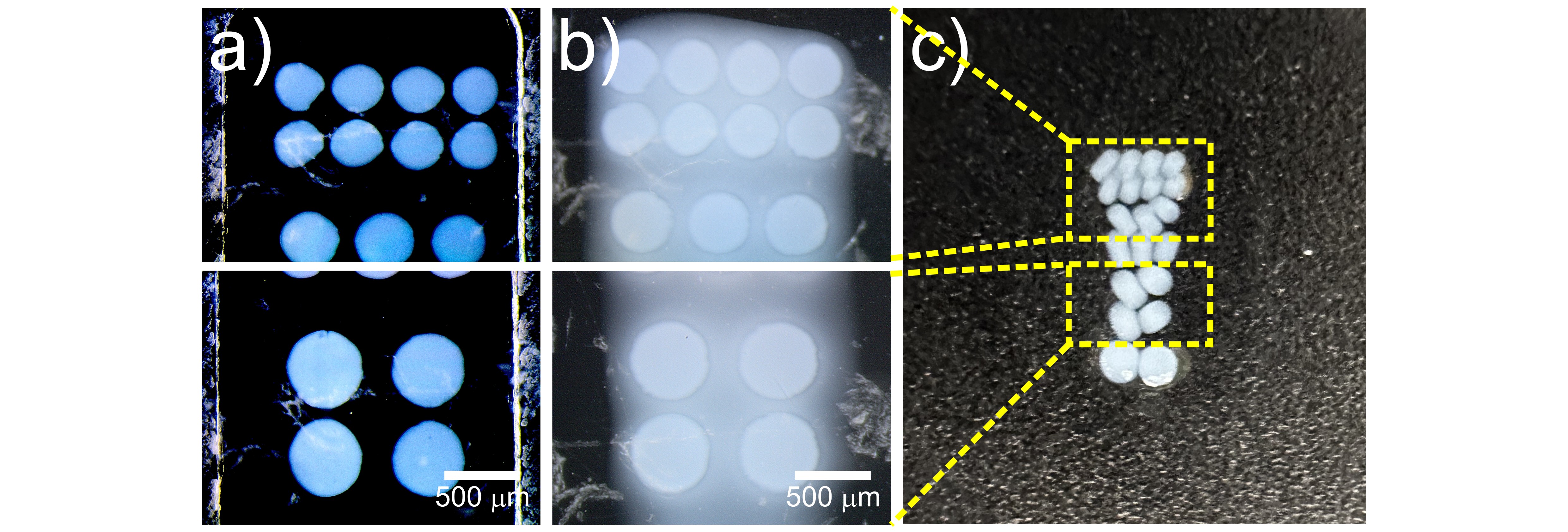}
 \caption{Stereo microscope images of (a) the hydrogel cylinders formed in PDMS channel, (b) the pattern formed in a bulk hydrogel strip through the Pt-coated mask. (c) Picture of hydrogel cylinders formed in a bulk hydrogel strip, which collapsed during washing steps.} 
  \label{SuppFig3}
\end{figure*}

\begin{figure*}
 \centering
 \includegraphics[width=1\textwidth]{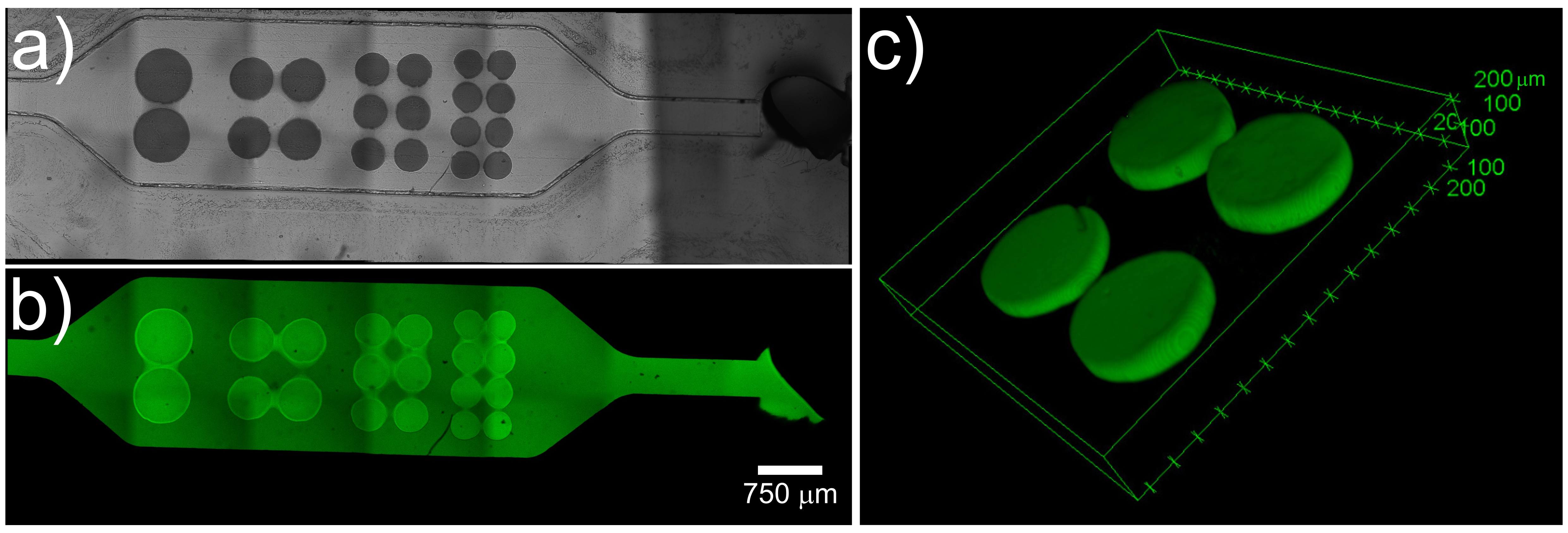}
 \caption{In situ and 3D formation of hydrogel cylinders within microchannels, shown with diffusion of fluorescent molecules. (a) Brightfield and (b) LSCM images of diffusion of 1 \textmu M Rhodamine 110 through an array of hydrogel cylinders in PDMS channel. (c) LSCM image with 3D view of hydrogel cylinders with Epirubicin hydrochloride diffusion.} 
  \label{SuppFig4}
\end{figure*}

\vskip3pt

\end{document}